\documentstyle[12pt,aasms4]{article}

\journalid{}{}
\articleid{}{}

\begin{document}

\title{A SEQUENCE OF DECLINING OUTBURSTS FROM GX339-4}

\received{25 June 1997}
\accepted{25 June 1997}

\author{B.C. RUBIN\altaffilmark{1,2}, B.A. HARMON\altaffilmark{3},
W.S. PACIESAS\altaffilmark{2,3}, C.R. ROBINSON\altaffilmark{3,4}, 
S.N. ZHANG\altaffilmark{3,4}, G.J. FISHMAN\altaffilmark{3}}

\altaffiltext{1}{Cosmic Radiation Laboratory, Institute of Physical and
Chemical Research (RIKEN), Wako-shi, Saitama 351-01 Japan. 
e-mail: rubin@crab.riken.go.jp}
\altaffiltext{2}{Department of Physics, University of Alabama in Huntsville,
Huntsville, AL 35899} 
\altaffiltext{3}{NASA/Marshall Space Flight Center, Huntsville, AL 35812}
\altaffiltext{4}{Universities Space Research Association.}

\begin{abstract}

The flux and spectrum
of the black hole candidate GX339-4 
has been monitored by
the Burst and Transient Source Experiment (BATSE) on the 
Compton Gamma-ray Observatory (CGRO) 
since the observatory became operational in May 1991.
Between the summer of 1991 and the fall of 1996,
eight outbursts from GX339-4 were observed. 
The history of these outbursts is one of declining
fluence or total energy release, as well as a shortening of the 
time between outbursts. A rough linear correlation exists between 
the fluence emitted
during an outburst and the time elapsed between the end of the previous 
outburst and the beginning of the current one. The peak flux is also
roughly linearly correlated with outburst fluence. The lightcurves of the
earlier, more intense, outbursts (except for the second one) can be modeled 
by a fast exponential 
(time constant $\sim$ 10 days) followed by a slower exponential 
($\sim$ 100 days) on the rise and a fast exponential decay ($\sim$ 5 days) on
the fall. The later, weaker, outbursts are modeled with a single rising time 
constant ($\sim$ 20 days) and a longer decay on the fall ($\sim$ 50 days).
An exponential model gives a marginally
better fit than a power law to the rise/decay profiles.
GX339-4 is a unique source
in having more frequent outbursts than other low mass x-ray 
binary black hole candidates. These observations can be used to constrain
models of the behavior of the accretion disk surrounding the compact
object.

\end{abstract}


\section{INTRODUCTION} \label{intro}

Much effort has been made to study the x-ray source GX339-4
since its original discovery by
OSO-7 (Markert, et al.\ 1973\markcite{mark73}). GX339-4
is usually considered a black hole candidate (BHC) due to the similarity of
its x-ray spectral and timing states to dynamical BHC such as Cygnus X-1, and
to the lack of detection of pulsations or x-ray bursts.
However, there is considerable uncertainty
about the mass of the compact object, and GX339-4 is not at this
time a dynamical BHC (Callanan, et al.\ 1992\markcite{call92}). 
However, it is similar to the two established jet sources 
GRS1915+105 and GROJ1655-40, the latter a dynamical BHC, in exhibiting 
multiple hard x-ray outbursts. 
The recent report  of a weak radio jet in GX339-4 
(Fender, et al.\ 1997\markcite{fend97})
(the velocity of the jet is 
essentially unkown) is a hint that these three sources may be 
fundamentally similar (Zhang, et al. 1997\markcite{zhang97}).
The existence of the radio jet requires confirmation from future observations.

Most x-ray observations of GX339-4 have focused on determining which
of four x-ray spectral states (off,low,high,very high)
the source is in, and describing the properties of those states
(Motch, et al.\ 1985\markcite{motch85}, for a recent review see 
Tanaka and Lewin 1995\markcite{tl95}).
Here we will present new data from six recent outbursts observed by
CGRO-BATSE. Our analysis will also include two earlier outbursts observed 
by BATSE (Harmon et al.\ 1994\markcite{harm94}).
The eight outbursts observed by BATSE can be roughly divided into two types
based on the outburst fluence, light curve, spectral evolution, and  
recurrence pattern. In these respects, the first
four outbursts appear to be different from 
the last four.

More important than this classification, however, is the observation that
the general pattern of these outbursts is one of decreasing total energy
release, as subsequent outbursts occur closer together in time, 
so that outburst fluence in the 20-300 keV band is correlated 
with the time elapsed since the previous outburst. 

\section{OBSERVATIONS} \label{observ}

Figure 1 shows the BATSE flux history obtained using the Earth occultation
technique and an optically thin thermal bremsstrahlung (OTTB) spectral model 
fit to the observed count rates in the 20-300 keV band. The functional form
of the OTTB model used is $A\exp(-E/kT)/E$ where E is the photon energy in keV
and the amplitude A and temperature $kT$ (in keV) are determined from the fit.
The observational techniques are described in
(Harmon et al.\ 1994\markcite{harm94}).

Figure 1 shows eight outbursts 
separated by intervals during which the source was not detected above the
$\sim 30$ mCrab threshold for ten day integrations. We will label these
outbursts B1-B8. 
The first three outbursts have similar lightcurves and recur at an
approximately periodic interval of $\sim 450$ days.
However, neither the profiles nor recurrence intervals of the later outbursts,
especially the last four, show evidence of 
being related to the earlier outbursts. 
Information on outburst beginning and ending times, peak fluxes and times,
and total fluences appears in Table 1.

Three spectral models: OTTB, photon power law (PL), and Sunyaev-Titarchuk
comptonization (ST) (Sunyaev \& Titarchuk 1980\markcite{st80})
were fit to the data over 20-300 keV.
An OSSE observation near the peak of B1 was consistent with an OTTB
model (kT $\simeq$ 70 keV)
over the full energy range in which the source was detected (up to 400 keV),
but inconsistent with PL and a marginal fit at best to ST above 200 keV 
(Grabelsky et al.\ 1995\markcite{grab95}). During almost all of B1-B4,
PL gives unacceptable fits to the BATSE data,
while the OTTB and ST models are both adequate, and fit equally well. 
Table 2 shows
a comparison of PL and OTTB model fits during selected intervals. 
While an OTTB model also always works during B5-B8, there are also
times during these outbursts when PL gives equally good fits.
During these times, the possibility that the spectrum is a power law which
also extends to higher energies can not be ruled out, though it is also
possible that an OTTB spectrum is always correct.

The spectral evolution during each outburst is presented in a plot
of OTTB fit temperature versus time
in Figure 2. During outbursts B1-B4
the temperature peaked early and declined gradually.
In contrast, it remains roughly constant during B5-B8.

\section{ANALYSIS OF THE OUTBURST PATTERN} \label{patt}

We have used the information in Table 1 
to examine the possibility of correlations among the outburst fluence, 
peak flux, duration, and time between outbursts.

The left panel of
Figure 3 shows that peak flux is roughly linearly correlated with
the total outburst fluence, over a factor of 3.5 in each parameter,
as first reported in Robinson et al. 1996\markcite{rob96}. 
Peak outburst luminosities depend
on source distance, which has been estimated 
between 1.3 kpc (Predehl et al. 1991\markcite{pred91}) 
and 4 kpc (Cowley et al. 1987\markcite{cow87}).
The peak luminosities range from 
$2.2\times 10^{35}d_{kpc}^{2} \rm{ergs/s}$ for B5
to $7.6\times 10^{35}d_{kpc}^{2} \rm{ergs/s}$ 
for B3 where $d_{kpc}$ is the source distance in kpc.

In the right panel of
Figure 3 the time elapsed since the previous outburst 
($T_{pi}$, the time between the end of the
previous burst B(i-1) and the start time of Bi) 
is plotted versus outburst fluence. 
An approximate linear correlation 
between $T_{pi}$ and fluence is observed, with the possible exception
of B5, which appears underluminous for this relation.
If we consider instead the time until the next outburst,
the deviations of B1 and
B3 from the trend are quite large. Thus, taking the time since the previous
outburst as the underlying
variable correlated with outburst energy release is the better description of
source behavior. This correlation implies that the time averaged luminosity,
$\overline{L} = 1.6\times 10^{35}d_{kpc}^{2} \rm{ergs/s}$, is roughly constant.
Outburst durations show no clear trend when plotted against fluence.

\section{OUTBURST TIMESCALES} \label{tmscale} 

Spectral fits in which flux is the only free parameter were used to
obtain a flux estimate for each day of data. 
In each fit, the temperature is held
fixed at the temperature determined from the corresponding ten day spectral
fit. We have attempted to model the first four outbursts 
(except for the second one) with 
an initial fast exponential rise followed by a second, much slower, 
rise. The last four outbursts (and the second one) have only a single rise,
and each outburst has a single decay.

Figure 4 shows the
one day resolution lightcurves of B1 and B5. 
Information about
rise and decay times of all of the outbursts is given in Table 3. In each
case the rise (including variable break time) and decay intervals 
were fit separately. 

For comparison we have also fit power law models to each rise/decay portion
of the lightcurve. F-test probabilities that the power law model is preferred
range from 3\% to 85\% and are typically about 30\%. Only on the fall of B1 
and the rise of B3 is this probability above 50\%. Exponential models are thus
marginally preferred.

\section{DISCUSSION} \label{discuss} 

In the past GX339-4 has been mostly observed sporadically by 
pointed instruments, 
making it difficult to discern any long term outburst pattern. 
Our analysis suggests that during the observations presented
here, there is a pattern in hard x-rays consisting of a 
sequence containing two types of outburst with declining fluence in the
20-300 keV band.

Nearly continuous observations made with the Ginga All Sky Monitor (ASM) 
in the 1-20 keV band
between early 1987 and the fall of 1991 indicate that this hard x-ray pattern 
may also be followed in soft x-rays
(Kitamoto, S.\ 1992\markcite{kita92}). 
During this time Ginga observed
three outbursts, which also follow a pattern of declining fluence, and which
could be an earlier part of the sequence observed by BATSE.
The first and brightest outburst occured after a long quiet period 
($>$ 1.5 years). 
During this outburst, the very high state was observed, probably
the only time it has been observed in this source.
The third outburst, which occurred near the end of the operational life of
Ginga and was only partially observed, was coincident with B1. 
A hard to soft transition occured about
$\sim 50$ days into this outburst, but there was otherwise no unusual soft
x-ray activity. These observations suggest the possibility that
the first outburst observed by Ginga initiated the subsequent
declining sequence.  However determining, whether, and to what extent, 
the patterns observed here in the BATSE data are also relevant below 20 keV, 
and in particular, the effect of x-ray spectral state changes,
will require more careful analysis of existing multi-instrument data.

For a constant mass accretion rate from the companion into an accretion disk, 
the correlation between fluence and recurrence time found here implies that
just the excess mass which accumulates in the disk between outbursts falls 
into the compact object during the outbursts. The rise and decay profiles
of the outbursts are related to how the matter passes through the disks,
and therefore to the properties of nonstationary accretion disks.
For example, exponential timescales
in nonstationary disks have been shown to imply a linear relation between
the diffusion coefficient and surface density; a non-linear relation would
imply power law profiles (Liubarskii and Shakura 1987\markcite{ls87}).   

A thermal instability in an outer thin disk can account for the recurrence,
rise, and decay times of soft x-ray transients (SXT) 
(Mineshige 1996\markcite{mine96}).
The outburst recurrence time is identified with the viscous timescale
in the "cool" (low $\alpha$) branch of the outer disk and the decay timescale
with the "hot" branch ($\alpha$ is the dimesionless viscosity). Thus, 
$t_{vis}=(R/H)^{2}(\alpha\Omega)^{-1} \simeq 10\sqrt{mR_{1}}(\alpha T)^{-1}$ 
where $m$ is the compact object mass in solar 
masses, $R_{1}$ the disk radius in $10^{10}$ cm and $T$ the temperature
in $10^{4}$ K. "Cold" $\alpha \sim 0.001-0.01$ implies 
a recurrence time of one to several years and 
"hot" $\alpha \sim 0.1-1.0$
a decay time of 10-100 days. The rise time is the sound speed
propagation time in the outer disk: 
$t_{rise} = R(\alpha C_{s})^{-1} \simeq 0.1R_{1}(\alpha T)^{-1/2}$,
on the order of a day for $\alpha \sim 0.1$.

Can similar arguments be applied to the case of GX339-4? 
Perhaps
only the fast decay times can not be explained in this way. These might instead
imply a shrinking of the thin disk radius, R1, by a factor of at least 10 or
more during the outburst. Shrinking may also suggest an explanation for the
dual risetimes in the early outbursts. If these outbursts originate deep inside
the thin disk, then the fast risetime represents the propagation
of the heating front throughout the disk, while the slow risetime is the 
viscous timescale on which the hot outer disk shrinks, as it pushes matter
into the inner disk. In this scenario, the first four
outbursts are "inside-out" with a shrinking disk while the last four, with
behavior closer to SXT, may be "outside-in". Time differences between the
origin of optical and x-ray radiation at the beginning of an outburst
can discriminate the direction of the outburst. For
example, an outburst of GROJ1655-40 was seen to be outside-in 
(Orosz et al 1997\markcite{orosz}). 
Such observations are strongly encouraged.

\acknowledgements

We acknowledge useful discussions with Dr. Konstantin Postnov.

\clearpage

\begin{figure}
\caption{BATSE lightcurve.
Photon flux as a function of time in the 20-300 keV energy band 
as determined from an optically thin thermal bremmstrahlung fit to Earth 
occultation data. 
Horizontal bars indicate the
data integration interval.
The vertical error bars are statistical only. 
Additional systematic uncertainties can arise from faint sources causing
interference along occultation edges. 
The data is histogrammed when an outburst is in progress. 
The numbers near the bottom of the plot label each of the outbursts B1-B8
(the B is omitted on the plot).
The dotted line is at the level of zero flux. 
} \label{figlc}
\end{figure}

\begin{figure}
\caption{Spectral Evolution of the outbursts. The spectral
evolution of each outburst is shown as a plot of OTTB model temperature
in the 20-300 keV band as a function of time since the beginning of the
outburst. The plot symbol corresponding
to each outburst is shown in the insets. 
Horizontal bars indicate data integration intervals.
The vertical error bars are statistical only.
The starting times of the outbursts are listed in Table 1
under the column heading 'Beginning Time'. 
} \label{figspecevol}
\end{figure}

\begin{figure}
\caption{Outburst parameters vs. outburst fluence in the 
20-300 keV band. In each of these plots, the number near each data point
labels the corresponding outburst (B1-B8). 
See Tables 1 and 3 for the values used in the plots.
Left panel: Outburst fluence versus peak flux.
Error bars are statistical only.
The dotted line shows the best fit straight line to the plotted data. 
Right panel: Outburst fluence versus time to the previous ourburst $T_{pi}$.
Vertical error bars are statistical only and horizontal bars show an
approximate uncertainty of 10 days in $T_{pi}$.
The dashed line is the best fit straight line. 
} \label{figcorr}
\end{figure}

\begin{figure}
\caption{One day resolution outburst lightcurves and model fits.
Top: B1. Time zero on the horizontal scale corresponds to Julian day 2448430.
Bottom: B5. Time zero corresponds to JD 2449840.
Vertical error bars are statistical only and horizontal
bars show the data integration interval.
The solid lines show the exponential rise and decay models discussed in the 
text.
The rise and decay timescales obtained from the fits are listed in Table 3. 
} \label{figtmscale}
\end{figure}

\clearpage

\begin{table*}
\begin{center}
\begin{tabular}{|c|c|c|c|c|c|}
\tableline
Outburst & Beginning Time & Ending Time & Peak Time & Peak Flux & Fluence \\
         & JD-2440000     & JD-2440000  & JD-2440000 & $\rm{photons/cm^{2}s}$ &
$\rm{ergs/cm^{2}}$ \\
\tableline
B1 & 8437 & 8537 & 8505 & 0.107 $\pm$ 0.002 & 0.038  \\
B2 & 8887 & 8982 & 8958 & 0.086 $\pm$ 0.002 & 0.048 \\
B3 & 9345 & 9438 & 9398 & 0.106 $\pm$ 0.001 & 0.054 \\
B4 & 9620 & 9691 & 9658 & 0.077 $\pm$ 0.001 & 0.032 \\
B5 & 9851 & 9937 & 9865 & 0.040 $\pm$ 0.001 & 0.014 \\
B6 & 9956 & 10025 & 9956 & 0.036 $\pm$ 0.002 & 0.016 \\
B7 & 10107 & 10168 & 10126 & 0.043 $\pm$ 0.001 & 0.016 \\
B8 & 10268 & 10347 & 10288 & 0.043 $\pm$ 0.001 & 0.023 \\
\tableline
\end{tabular}
\end{center}

\caption{Outburst Beginning, Ending, and Peak Times, Fluxes, and Total 
Fluences.
Peak fluxes in each outburst are calculated as the largest average of
three consecutive time integrations. 
Peak times are at the center of these averages.
}\label{tablist1}
\end{table*}

\clearpage

\begin{table*}
\begin{center}
\begin{tabular}{|c|c|c|c|c|c|c|}
\tableline
Outburst & Times & OTTB Temperature & $\chi^{2}_{OTTB}$ & PL index & 
$\chi^{2}_{PL}$ & $\nu$ \\
\tableline
B1 (R2) & 8504-8511   & 61 $\pm$ 3  & 20  & 2.4  & 94   & 16  \\
B2 (M)  & 8950-8957   & 78 $\pm$ 5  & 33  & 2.2  & 65   & 25  \\
B3 (R2) & 9368-9384   & 70 $\pm$ 3  & 51  & 2.2  & 108  & 34  \\
B4 (R2) & 9643-9657   & 65 $\pm$ 3  & 59  & 2.2  & 79   & 34  \\
B5 (M)  & 9874-9898   & 86 $\pm$ 7  & 93  & 2.1  & 93   & 70  \\
B6 (R)  & 9951-9967   & 77 $\pm$ 9  & 16  & 2.2  & 22   & 16  \\
B7 (M)  & 10126-10147 & 87 $\pm$ 6  & 45  & 2.0  & 79   & 46  \\
B8 (D)  & 10308-10322 & 96 $\pm$ 22 & 41  & 2.0  & 47   & 32  \\
\tableline
\end{tabular}
\end{center}

\caption{OTTB and Power Law (PL) Spectral Fits in the 20-300 keV range.
The portion of the outburst from which the data is drawn is given in
parentheses in the Outburst column. R = Rise; R2 = Second rise; M = Middle;
D = Decay. The times are in JD-2440000.5 (Truncated Julian Days). Uncertainties
in the power law index are 0.1 or less. $\nu$ is the number of degrees of
freedom in each of the spectral fits. Nine channels from each detector in each
spacecraft pointing interval were used. Most of the fits encompass multiple
pointing intervals.
}\label{tablist2}
\end{table*}

\clearpage

\begin{table*}
\begin{center}
\begin{tabular}{|c|c|c|c|c|c|c|c|}
\tableline
Outburst & Previous & Duration & Rise 1 & Rise 2 & $\chi^{2}_{\nu rise} \,
(\nu)$ & Decay & $\chi^{2}_{\nu decay} \, (\nu)$ \\
\tableline
B1 &   - & 100 & 12.7 $\pm$ 1.5 & 153 $\pm$ 52 & 1.6 (65) & 5.7 $\pm$ 0.5 
& 1.5 (5) \\
B2 & 350 &  95 &  -             & 108 $\pm$ 25 & 0.9 (39) & 3.4 $\pm$ 0.5 
& 0.8 (8) \\
B3 & 363 &  93 &  7.6 $\pm$ 1.5 & 104 $\pm$ 12 & 3.8 (73) & 8.9 $\pm$ 0.9 
& 0.6 (10) \\
B4 & 182 &  71 &  21 $\pm$ 11   &  56 $\pm$ 12 & 1.4 (37) & 5.8 $\pm$ 0.8 
& 0.5 (6) \\
B5 & 160 &  86 &  19 $\pm$ 13   &  -           & 0.8 (8)  & 56 $\pm$ 14 
& 1.0 (36) \\
B6 &  19 &  69 &  18 $\pm$  7   &  -           & 1.0 (8)  & 55 $\pm$ 24 
& 1.3 (22) \\
B7 &  82 &  61 &  13.5 $\pm$ 2.6 &  -          & 0.7 (16) & 12.6 $\pm$ 3.6 
& 0.8 (11) \\
B8 &  99 &  79 &  47 $\pm$  6   &  -           & 1.3 (32) & 64 $\pm$ 18 
& 0.7 (13) \\
\tableline
\end{tabular}
\end{center}

\caption{Timescales During Outburst Sequence.
The second column (labeled Previous) is $T_{pi}$,
the time between the end of the previous and the begininning of the
current outburst. Duration is the duration of the outburst, given by the
difference between the beginning and ending times in Table 1. 
Rise 1 is the first e-folding rise time and Rise 2 the second
e-folding rise time determined from an exponential model fit (with variable
break time, when both timescales are present) on the rising portion of
the outburst. Decay is the e-folding decay time from an exponential model fit 
on the falling side of the outburst. 
All times are in days. The reduced $\chi^{2}$ and corresponding
number of degrees of freedom, $\nu$, are shown for each fit.
}\label{tablist3}
\end{table*}


\clearpage

\begin{references}

\reference{call92}
Callanan, P. J., Charles, P. A., Honey, W. B., Thorstensen, J. R. 1992,
\mnras, 259, 395

\reference{cow87}
Cowley, A. P., Crampton, D., \& Hutchings, J. B. 1987, AJ, 92, 195

\reference{fend97}
Fender, R. P., Spencer, R. E., Newell, S. J., Tzioumis, A. K. 1997,
\mnras, 286, L29

\reference{grab95}
Grabelsky, D. A., Matz, S. M., Purcell, W. R., Ulmer, M. P.,
Grove, J. E., Johnson, W. N., Kinzer, R. L., Kurfess, J. D., 
Strickman, M. S. 1995, \apj, 441, 800

\reference{harm94}
Harmon, B. A., Wilson, C. A., Paciesas, W. S., Pendleton, G. N.,
Briggs, M. S., Rubin, B. C., Finger, M. H., Fishman, G. J., et al. 1994,
\apj, 425, L17

\reference{kita92}
Kitamoto, S. in ``Proceediings of the 4th International Conference on Plasma
Physics and Controlled Nuclear Fusion'', 1992, 297

\reference{ls87}
Liubarskii, Y. E. \& Shakura, N. 1987, Soviet Astronomy Letters, 13, 917

\reference{mark73}
Markert, T. H., Canizares, C. R., Clark, G. W., Lewin, W. H. G.,
Schnopper, H. W., \& Sprott, G. F. 1973, \apj, 184, L67

\reference{mine96}
Mineshige, S. 1996, PASJ, 48, 93.

\reference{motch85}
Motch, C., Ilovaisky, S. A., Chevalier, C., Angebault, P. 1985,
\ssr, 40, 219

\reference{orosz97}
Orosz, J. A., Remillard, R. A., Bailyn, C. D., McClintock, J. E. 
1997, \apj, 478, L83 

\reference{pred91}
Predehl, P., Brauninger, H., Burkert, W., \& Schmitt, J.. et al. 1991,
A \& A, 246, L40

\reference{rob96} Robinson C. R., et al. 1996, in {\it Proceedings of the
Second Integral Workshop 'The Transparent Universe'}, ESA SP-382, 249

\reference{st80}
Sunyaev, R. A., \& Titarchuk, L. G. 1980, A \& A, 86, 121

\reference{tl95}
Tanaka, Y., Lewin, W., in ``X-Ray Binaries'',
Lewin, W., van Paradijs, J., van den Heuvel, E., editors,
Cambridge University Press, 1995, p. 126.

\reference{zhang97}
Zhang S. N., Mirabel, I. F., Harmon, B. A., Kroeger, R. A., 
Rodriguez, L. F., Hjellming, R. M., Rupen, M. P.,
to appear in ``Proceedings of the 4th Compton Symposium'',
American Institute of Physics, Williamsburg Virginia, April 1997

\end{references}
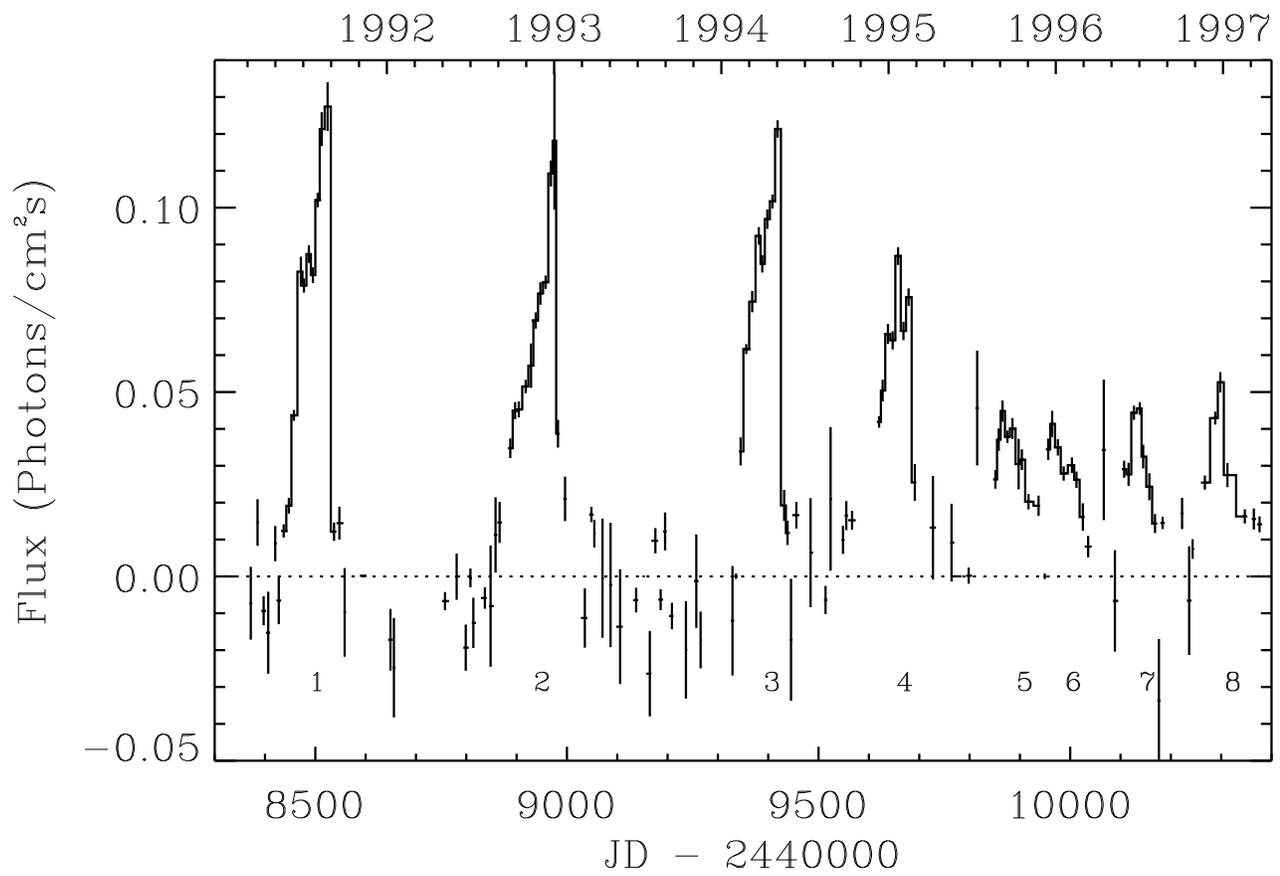
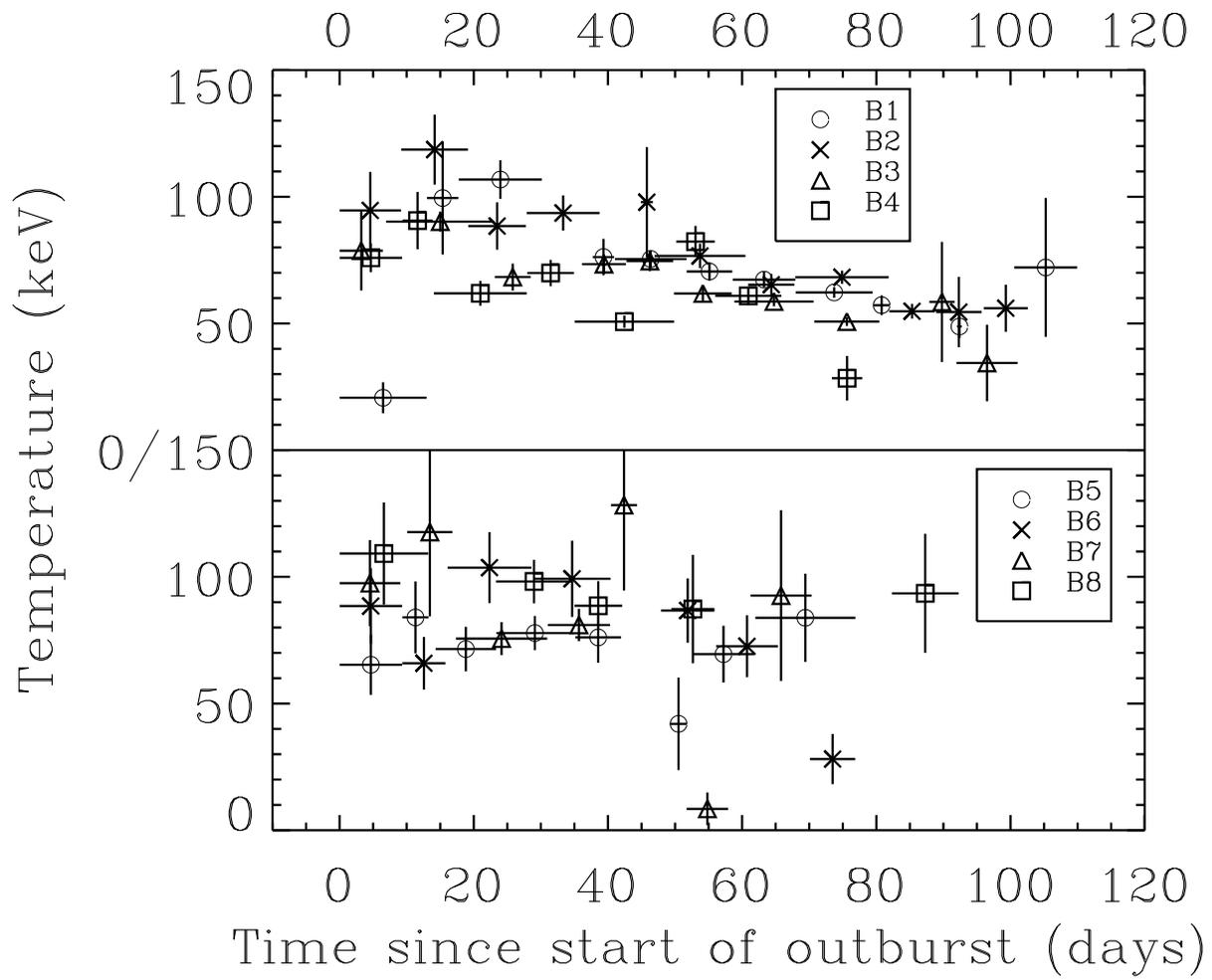
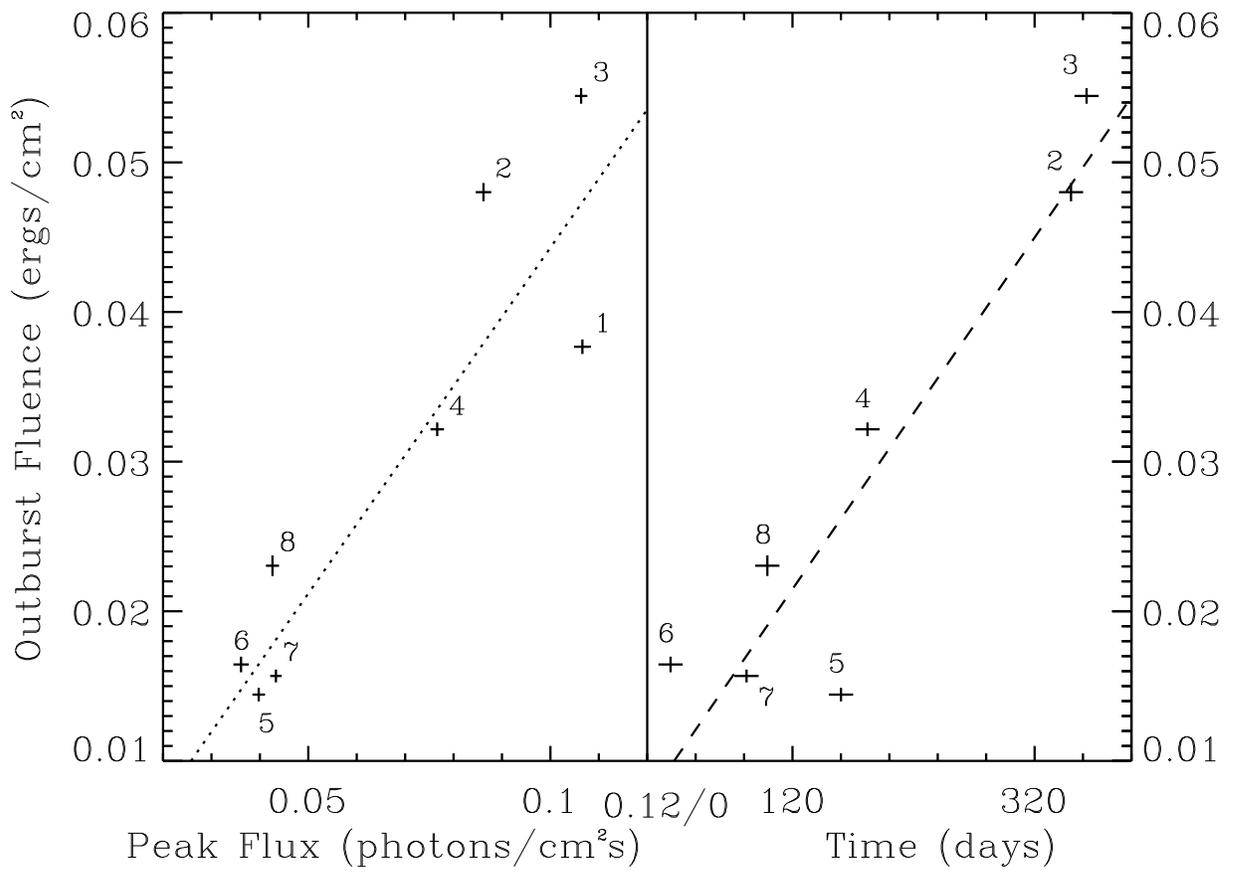
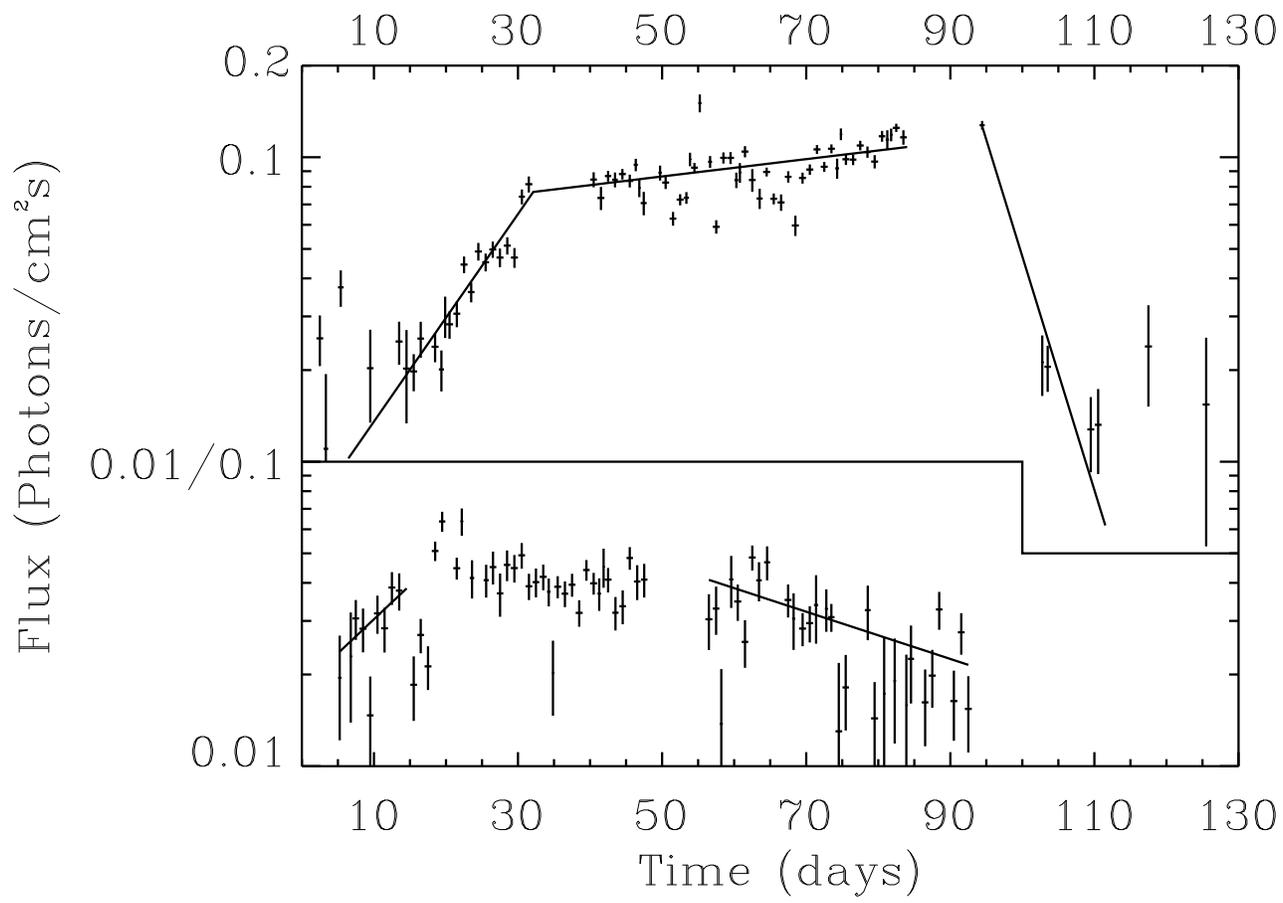
\end{document}